\input harvmac
\noblackbox
\def\Title#1#2{\rightline{#1}\ifx\answ\bigans\nopagenumbers\pageno0\vskip1in
\else\pageno1\vskip.8in\fi \centerline{\titlefont #2}\vskip .5in}

\def\msurr{\mathsurround=0pt}
\def\overleftrightarrow#1{\vbox{\msurr\ialign{##\crcr
	$\leftrightarrow$\crcr\noalign{\kern-1pt\nointerlineskip}
	$\hfil\displaystyle{#1}\hfil$\crcr}}}
\def\lrd{{\overleftrightarrow \partial}}
\def\ads{anti-de Sitter}
\def\gbdel{{G_{B\partial}}}
\def\sq{{\vbox {\hrule height 0.6pt\hbox{\vrule width 0.6pt\hskip 3pt
   \vbox{\vskip 6pt}\hskip 3pt \vrule width 0.6pt}\hrule height 0.6pt}}}
\def\vecm{{\vec m}}
\def\hf{{1\over 2}}
\def\poin{Poincar\'e}
\def\aolm{\alpha_{\omega l\vecm}}

\def\limrho{\buildrel \rho\rightarrow \pi/2 \over \longrightarrow}

\def\ylm{Y_{l\vecm}}
\def\eg{{\it e.g.}}
\def\rhob{{\bar \rho}}
\def\lrdm{\overleftrightarrow{\partial_\mu}}
\def\lrdr{\overleftrightarrow{\partial_\rho}}
\def\calo{{\cal O}}
\def\rin{{\rm in}}
\def\rout{{\rm out}}
%
%
\def\ajou#1&#2(#3){\ \sl#1\bf#2\rm(19#3)}
\def\jou#1&#2(#3){,\ \sl#1\bf#2\rm(19#3)}
%
%

\font\ticp=cmcsc10
%
%
\lref\GKP{S.S. Gubser, I. Klebanov and A. Polyakov, ``Gauge
theory correlators from noncritical string theory'', hep-th/9802109,
\ajou Phys. Lett. &B428 (98) 105.}

\lref\Mald{J. Maldacena, ``The large-N limit of superconformal
field theories and supergravity'', hep-th/9711200, \ajou
Adv. Theor. Math. Phys. &2 (98) 231.}

\lref\Witt{E. Witten, ``Anti-de Sitter space and holography'',
hep-th/9802150, \ajou Adv. Theor. Math. Phys. &2 (98) 253.}

\lref\BKL{V. Balasubramanian, P. Kraus and A. Lawrence,
``Bulk vs. Boundary dynamics in anti-de Sitter spacetime'',
hep-th/9805171, \ajou Phys. Rev & D59:046003 (99).}

\lref\BKLT{V. Balasubramanian, P. Kraus, A. Lawrence and S. Trivedi,
``Holographic probes of anti-de Sitter space-times'',
hep-th/9808017, Harvard preprint HUTP-98/A057, Caltech preprint
CALT68-2189, Fermilab preprint Pub-98/240-T}

\lref\BDHM{T. Banks, M. Douglas, G. Horowitz and E. Martinec,
``AdS dynamics from conformal field theory'', hep-th/9808016,
UCSB/ITP preprint NSF-ITP-98082, U. Chicago preprint EFI-98-30.} 

\lref\Polc{J. Polchinski, ``S-matrices from AdS spacetime'',
hep-th/9901076.}

\lref\Suss{L. Susskind, ``Holography in the flat-space 
limit'', hep-th/9901079.}

\lref\DuFr{D.W. D\"usedau and D.Z. Freedman,
``Lehmann spectral representation for anti-de
Sitter quantum field theory'', \ajou Phys. Rev. &D33 (86) 389.}

\lref\Araf{I.Ya. Aref'eva, ``On the holographic S-matrix,'' hep-th/9902106.}

\lref\PoSu{J. Polchinski and L. Susskind, ``Puzzles and paradoxes about
holography,'' hep-th/9902182.}

\lref\HoIt{G. Horowitz and N. Itzhaki, ``Black holes, shock waves, and
causality in the AdS/CFT correspondence,'' hep-th/9901012.}

\lref\BGL{V. Balasubramanian, S.B. Giddings, and A. Lawrence, ``What do
CFTs 
tell us about anti-de Sitter spacetimes?'' hep-th/9902052.}

\Title{\vbox{\baselineskip12pt\hbox{hep-th/9903048}
}}
{\vbox{\centerline {The boundary S-matrix}
\vskip2pt\centerline{and the AdS to CFT dictionary}
}}
\centerline{{\ticp Steven B. Giddings}\footnote{$^\dagger$}
{Email address:
giddings@physics.ucsb.edu}
}
\vskip.1in
\centerline{\sl Department of Physics}
\centerline{\sl University of California}
\centerline{\sl Santa Barbara, CA 93106-9530}

\bigskip
\centerline{\bf Abstract}

An S-matrix analog is defined for anti-de Sitter space by constructing
``in'' and ``out'' states that asymptote to the timelike boundary.  A
derivation parallel to that of the LSZ formula shows that this ``boundary
S-matrix'' is given directly by correlation functions in the boundary
conformal theory.  This provides a key entry in the AdS to CFT dictionary.

\Date{}

The conjectured AdS/CFT correspondence\Mald\ has offered a promising new
window into the dynamics of string/M theory.  But in order to exploit this
powerful framework, we must decipher the holographic relationship between
the bulk and boundary theories.  Gubser, Klebanov, and Polyakov\GKP\ and
Witten\Witt\ made important progress in this regard by providing a CFT to
AdS dictionary: they show how to derive CFT correlation functions from the bulk
theory in AdS.  This has allowed the successful calculation of various CFT
correlators.  

However, in order to study bulk physics, and in particular to understand
the undoubtedly profound implications of holography, a reverse dictionary
is needed: we need to know which bulk 
quantities can be calculated, and how to calculate them, from the boundary CFT.
Another important and closely related question is how to
treat scattering in AdS.  Due to the periodicity of particle orbits and
lack of ordinary asymptotic states in AdS, a conventional S-matrix cannot
be defined.\foot{See \refs{\BGL} for more discussion.}  However, \BGL\
outlined the definition of an AdS analog of the S-matrix in terms of
scattering of states from the timelike 
infinity.  Refs.~\refs{\Polc,\Suss} gave a
related definition in the infinite-$N$ limit.  The purpose of the present
note will be to go further and provide an intrinsic and explicit 
definition of this ``boundary''
S-matrix for arbitrary $N$, and to give a precise relation between it and the CFT
correlators.  The
discussion also clarifies the relation between the framework of
\refs{\Polc,\Suss} and that of \BGL.  Other recent treatments of related
aspects of the AdS/CFT dictionary include\refs{\HoIt,\Araf,\PoSu}.

To summarize in advance, the boundary S-matrix will be defined as an
overlap of certain ``in'' and ``out''
states.  These will be defined so that they correspond to particles
asymptotic to the timelike boundary of AdS in the  past/future.  An
AdS analog of the LSZ formula can then be derived and relates this S-matrix to
the bulk correlation functions.  Finally the results of \Witt\ are used
to rewrite the boundary S-matrix in terms of the CFT correlation
functions.  An extremely simple relationship results:  the boundary
S-matrix {\it equals} the corresponding CFT correlator.  
This serves as a key entry in the AdS to CFT dictionary.

For simplicity we will consider scalar fields, with action
\eqn\action{S=-\int dV \left[ \hf (\nabla \Phi)^2 + {m^2\over 2}
\Phi^2+ U(\Phi)\right]\ ,}
where $U$ summarizes the interaction terms.  The generalization to other
fields should be straightforward.  
We will work in global coordinates $x=(t,\rho,\Omega)$ for $AdS_{d+1}$,
\eqn\glocoor{ds^2 = R^2 ( -\sec^2\rho \, dt^2 + \sec^2\rho \, d\rho^2 +
\tan^2\rho \, d\Omega^2_{d-1} )\ ,
}
although translation to \poin\ coordinates should also be
straightforward.

Certain facts about the solutions to the free equations will be useful in the
following.  
The effective gravitational potential of \ads\ space confines 
particles to its interior.  Solutions to the free equation
\eqn\scalsoln{(\sq - m^2) \phi =0}
therefore exist at arbitrary frequency $\omega$, but are only
normalizable (in the Klein-Gordon norm) for a discrete set of frequencies.
Define the parameters $h_\pm$ and $\nu$ by
\eqn\hndef{2h_\pm = {d\over 2} \pm \nu\ ;\ \nu=\hf\sqrt{d^2+4m^2R^2}\ .}
Normalizable solutions with definite angular momenta are of the form
\eqn\normsoln{\phi_{nl\vecm} = e^{-i\omega_{nl} t} \ylm \chi_{nl}(\rho)}
and have asymptotic behavior 
\eqn\normasym{\chi_{nl}(\rho)\limrho k_{nl} (\cos\rho)^{2h_+} }
for some constants $k_{nl}$.  Explicit formulas for these solutions are
given in \refs{\BKL}.  The discrete eigenfrequencies are 
\eqn\efreq{\omega_{nl} = 2 h_+ +2n+l\ ,\ n=0,1,2,\cdots\ .}
Non-normalizable solutions we will write as
\eqn\nonnorm{\phi_{\omega l\vecm} = e^{-i\omega t} \ylm \chi_{\omega
l}(\rho)\ .}
These have asymptotic behavior
\eqn\nnormas{\chi_{\omega l}\limrho (\cos \rho)^{2h_-}\ ,}
where a convenient normalization convention has been chosen by fixing the
overall constant.

We will also require some assumptions about the spectrum.  This is classified
according to the representations of $SO(d,2)$.  A given representation is
determined by its weight $\Delta$ (which corresponds to the conformal
weight in the CFT), and contains states $|\Delta;n,l,\vecm\rangle$.  Here
$n$ is the principal quantum number and $l,\vecm$ are the standard angular
quantum numbers, as above.  
The energy, defined with respect to
global time, is given by
\eqn\adsener{\omega^\Delta_{n\ell} = \Delta +2n + l\ .}
For a free field, 
\eqn\freed{\Delta=2 h_+\ .}

We assume that the states of the interacting scalar theory consist of the
vacuum, $|0\rangle$, the single particle states,
$|\Delta;n,l,\vecm\rangle$, and multiparticle states 
$|\Delta_\beta;n,l,\vecm\rangle_\beta$ where $\beta$ is an additional state
label.  For the interacting field $\Delta$ may be renormalized and 
is not necessarily given in terms of the bare mass by  \freed.  
We also assume that 
\eqn\specass{\Delta_\beta> \Delta}
for all multi-particle states.

Some useful
properties of the bulk-boundary propagator are also needed.
Suppose that we
seek a solution of the free equation \scalsoln\ 
satisfying the boundary condition 
\eqn\phibc{\phi\limrho (\cos\rho)^{2h_-} f(b)\ ,}
where $b=(t,\Omega)$ denotes the boundary
coordinates and $f$ is some specified boundary value. 
Witten\refs{\Witt} defines the bulk-boundary Green function to be the
kernel that provides the solution:\foot{Though Witten's definition was made
in the euclidean continuation of AdS, the formalism naturally extends to
lorentzian signature as
discussed in \refs{\BKLT,\BGL}.}
\eqn\bbdefone{\phi(x) = \int db f(b) \gbdel(b,x)\ .}
Explicit expressions for $\gbdel$ can then be found\Witt\
using the resulting
condition that $\gbdel$ must asymptote to a delta function at the boundary.

It is easy to derive an alternate formula for  $\gbdel$ in terms of the bulk
Feynman propagator $G_B(x,x')$ using an AdS variant of the usual Green's
theorem argument.  
Consider the solution
$\phi$ with the above boundary conditions \phibc.  Define a region $V$ 
by $\rho<\rhob \approx \pi/2$.  
Using
\eqn\gbdef{(\sq_x - m^2)G_B(x,x') = -\delta(x,x')}
we may rewrite $\phi$    as
\eqn\trivrel{\phi(x') = -\int_V dV\phi(x) (\sq_x - m^2)G_B(x,x') }
and then integrate twice by parts to find
\eqn\greenthm{\phi(x') = -\int_{\partial V} d\Sigma^{\mu} \phi(x) \lrdm G_B(x,x')\
.}
At the boundary 
the Feynman propagator scales as 
\eqn\gbscal{G_B(x,x')\limrho (\cos \rho)^{2 h_+} G(b,x')\ ,}
for some function $G$.
Substituting this and the boundary behavior \phibc\ into \greenthm\ 
gives 
\eqn\intermed{\phi(x') = -R^{d-1} \int db f(b) G(b,x')
\lim_{\rho\rightarrow \pi/2}(\tan\rho)^{d-1}
\left[(\cos\rho)^{2h_-}\lrdr (\cos\rho)^{2h_+}\right]\ ,}
with limit 
\eqn\phires{\phi(x') = 2\nu R^{d-1} \int db f(b) G(b,x')\ .}
Comparison with \bbdefone\ then shows
\eqn\gbbdef{\gbdel(b,x') = 2\nu R^{d-1}
\lim_{\rho\rightarrow\pi/2} (\cos\rho)^{-2h_+}
G_B(x,x')\ ,}
in agreement with \refs{\BDHM}.

Notice that \scalsoln, \phibc\ do not uniquely specify the solution
$\phi$; one may always add a normalizable mode without modifying the
boundary behavior \phibc.  Specifically, suppose that $f(b)$ falls to zero
in the far past and future.  Then in the asymptotic past and future $\phi$
must be a linear combination of the normalizable modes \normsoln.  
With the preceding construction of $\gbdel$,
\bbdefone\ gives the solution that is purely positive frequency
in the far
future, and purely negative frequency in the far past.
Other solutions can be
obtained by modifying the temporal boundary conditions on the bulk Green
function, \eg\ by using retarded or advanced propagators.

Note that the non-normalizable solutions $\phi_{\omega l\vecm}$ can be
recovered from \bbdefone; in the limit $f\rightarrow e^{-i\omega t}\ylm$,
\nnormas\ coincides with \phibc.  Therefore 
\eqn\nnbb{\phi_{\omega l\vecm}(x) = \int db e^{-i\omega t} \ylm \gbdel(b,x)
\equiv \gbdel(-\omega,l,-\vecm;x)\ .}

For a general $f$, we can therefore rewrite \bbdefone\ in terms of the
Fourier transform $f_{l\vecm}(\omega)$ as 
\eqn\wavepack{\phi_f(x) = \int db f(b) \gbdel (b,x)=
\sum_{l,\vecm} \int {d\omega\over 2\pi}
f_{l\vecm}(\omega) \phi_{\omega l\vecm}(x)\ }
For appropriately chosen $f$, this function defines a solution
corresponding to a wavepacket.  The function $f$ determines the
packet profile.  

These can be thought of as packets incident from infinity in
AdS.  Notice that, according to \nnormas, they typically diverge at the
boundary.  There is a physical reason for this:  motion in the region near
the boundary is classically forbidden.  Therefore 
the amplitude for a particle incident
from infinity to reach the center of AdS is suppressed by an infinite
tunneling factor.  However, the amplitude for a particle to reach the
center of AdS may be kept finite by rescaling the wavefunction such that
the incident amplitude at infinity is infinite.  

One concrete way to think
of this is to imagine cutting off AdS at large but finite radius and
patching the resulting AdS bubble into a spacetime with a bona-fide null
infinity, as in \BGL.  A beam of particles from this asymptotic space can
be focussed to collide with another beam in the center of the AdS region.
Most of the incident flux is reflected off the potential barrier resulting
from the AdS geometry, 
so in order for the beams to
penetrate to the center the incident amplitudes must be large.  The
wavepacket definitions above, which are given intrinsically in AdS
without reference to an auxiliary bubble picture, can be thought of as
arising from the limit where the radius of the AdS bubble goes to infinity
while simultaneously scaling up the incident beam amplitudes.

These wavepackets can now be used to construct operators that create ``in''
and ``out'' states.\foot{Refs.~\refs{\Polc,\Suss} 
outlined the construction of such operators in the
infinite-$N$ limit.  Here we will explicitly construct such operators for
arbitrary $N$.}
These asymptotic operators will be defined by
\eqn\indef{\alpha_f= \lim_{\rhob\rightarrow \pi/2}
\int_\Sigma d\Sigma^\mu \phi_f^* \lrd_\mu
\Phi\ ,}
where 
$\Sigma=\partial V$ for the region $V$ defined above and $\Phi$ is the
full {\it interacting} field.  We also define the plane wave limit of these
operators, 
\eqn\pllim{\alpha_{\omega l\vecm} =  \lim_{\rhob\rightarrow \pi/2}
\int_\Sigma d\Sigma^\mu 
\phi_{\omega l\vecm}^* \lrdm \Phi\ .}
If $\Phi$  is replaced by the
free field,
\eqn\freefd{\phi = \sum_{n,l,\vecm} a_{nl\vecm} e^{-i\omega_{nl} t}\phi_{nl\vecm}
+ a_{nl\vecm}^\dagger e^{i\omega_{nl} t} \phi_{nl\vecm}^*\ ,}
then \pllim\ gives
\eqn\freein{\aolm = -4\pi\nu R^{d-1} \sum_n k_{nl}\left[ \delta(\omega-\omega_{nl})
  a_{nl\vecm} + \delta(\omega+\omega_{nl})  a_{nl,-\vecm}^\dagger\right] \ }
where the $k_{nl}$ appeared in \normasym.
This suggests that the positive and negative frequency $\aolm$'s be thought
of as annihilation and creation operators, respectively.

This is confirmed by the following critical relations, which hold for the
operators constructed from the full interacting field:
\eqn\overlapa{\langle 0| \alpha_f |\Delta; n,l,\vecm\rangle =
-2\nu R^{d-1} N(\Delta) k_{nl} f_{l\vecm}^*(\omega_{nl})\ ,} 
%
\eqn\overlapb{ \langle \Delta; n,l,\vecm| \alpha_f |0\rangle
= -2 \nu R^{d-1} N(\Delta) k_{nl} f_{l,-\vecm}^*(-\omega_{nl})\ ,} 
%
and
\eqn\overlapc{ \langle 0| \alpha_f |\Delta_\beta;
n',l',\vecm'\rangle_\beta = 0\ ,}
where $N(\Delta)$ is another constant.
Therefore $\alpha_f$ with positive-frequency $f$ annihilates a particle at
the boundary,
and with negative-frequency $f$ creates a particle at the boundary.  Furthermore,
\overlapc\ implies that the $\alpha_f$'s only annihilate/create {\it
single} particle states.

The first of these relations is proved
by recalling that by symmetry, the
full interacting field must satisfy\refs{\DuFr}
\eqn\interfcn{\langle 0| \Phi(x) | \Delta; n,l,\vecm\rangle  = N(\Delta)
\phi_{nl\vecm}(x)\ }
for some normalization factor $N(\Delta)$.
Then the definition \pllim\ and a derivation like that in
\intermed-\phires\ immediately gives \overlapa.
Note that in order for this to be true the modes in \wavepack, \pllim\ must
be 
defined
with the mass fixed by $2h_+=\Delta$, corresponding to 
using the renormalized physical mass of the
single particle state.
Analogous reasoning proves \overlapb.

Eq.~\overlapc\ is shown by noting that, again purely from the $SO(d,2)$ symmetry,
\eqn\multio{\langle 0| \Phi(x) | \Delta_\beta; n,l,\vecm\rangle_\beta  = 
N_\beta(\Delta_\beta)\phi^{\Delta_\beta}_{nl\vecm}(x)\ ,}
where $\phi^{\Delta_\beta}_{nl\vecm}$ is defined using the mass parameter
corresponding to the multiparticle $\Delta_\beta$.  Again, the matrix
element \overlapc\
can
be found from reasoning parallel to \intermed-\phires, but now the result
contains 
\eqn\dvanish{\lim_{\rho\rightarrow\pi/2} (\cos\rho)^{\Delta_\beta - \Delta}\ .}
This vanishes by \specass.

``In'' and ``out'' states are now readily defined.  For positive-frequency
functions $f_i$, define
\eqn\inop{ \alpha_{f_i}^\rin = \alpha_{f_i}/Z_i\ ,}
and for negative-frequency functions $f_j'$ define 
\eqn\outop{ \alpha_{f_j'}^{\rout\dagger} = \alpha_{f_j'}/Z_j'\ }
where the $Z_i$ are wavefunction renormalization factors necessary to cancel
normalization  constants like those in \overlapa,\overlapb.
The ``in'' and ``out'' states are then
\eqn\instate{|f_i\rangle_{\rm in} = \prod_i \alpha_{f_i}^{\rin\dagger} |0\rangle}
and 
\eqn\outstate{{}_{\rout}\langle f_j' | = \langle 0| \prod_j
\alpha_{f_j'}^\rout\ .}
These states in turn lead to construction of the boundary S-matrix.  Suppose that the
wavepackets $f_i, f_j'$ are non-overlapping, and that the support of all
the $f_j'$'s lies to the future of that of all the $f_i$'s.  The boundary
S-matrix is then defined as 
\eqn\bsmat{S_\partial[f_1\cdots f_m;f_1'\cdots f_n'] \equiv \langle 0|T
\prod_j \alpha_{f_j'}^\rout \prod_i \alpha_{f_i}^{\rin \dagger} |0\rangle\ .}
Although the interpretation is less transparent, the same definition can be
adopted for $f_i$ and $f_j'$ not satisfying the above conditions.

In flat space, the  S-matrix is related by the LSZ formula to truncated
correlation functions.  A similar formula can now be derived for the
boundary S-matrix, which will be given in terms of bulk correlation
functions.  Consider a finite region $V'$ defined like $V$ above, but with
boundaries $\Sigma_{-T}, \Sigma_T$ of constant time $\pm T$ 
lying to the far past and
far future of the wavepackets' support.  Gauss' theorem applied to \indef\
gives 
\eqn\gaussthm{ \alpha_f^{\rin\dagger} 
={1\over  Z} \lim_{\rhob\rightarrow \pi/2}\left[ \int_{V'} dV
\nabla^\mu (\phi_f \lrdm \Phi) - \int_{\Sigma_{-T}+\Sigma_T} d\Sigma^\mu
\phi_f \lrdm \Phi\ \right]\ }
where the $T\rightarrow\infty$ limit is also understood.
The surface term at $\Sigma_{-T}$ vanishes because $\phi_f$ is positive
frequency and therefore vanishes in the far past.  Inside
\bsmat, the time ordering takes the surface term at $\Sigma_T$ to the
left.  We can then insert a complete set of states $|s\rangle$ 
 to find an expression of
the form 
\eqn\tovanish{  \sum_s \int_{\Sigma_T} d\Sigma^\mu \phi_f \lrdm \langle
0| \Phi |s\rangle \langle s|\psi\rangle \ .}
This vanishes by \interfcn.  The bulk term is left, and after using
the free equation for $\phi_f$ and taking $T\rightarrow\infty$ becomes
\eqn\alphaeq{\alpha_f^{\rin\dagger}\approx{1\over Z} 
\int dV \phi_f (\sq-m^2) \Phi }
where $\approx $ denotes equality inside \bsmat.  Similar arguments hold
for $\alpha_f^\rout$.  The LSZ formula
immediately follows:
\eqn\lsz{S_\partial[f_1\cdots f_m;f_1'\cdots f_n'] = \int
\prod_i\left[{dV_i\over Z_i}
\phi_{f_i}(x_i) \right] \prod_j\left[{dV_j'\over Z_j'}
\phi_{f_j'}(x_j') \right] \langle 0| T \prod_i\Phi(x_i)
\prod_j\Phi(x_j') |0\rangle_T \ .}
Here the kinetic operator in \alphaeq\ has removed the external legs,
and the ``T'' subscript denotes the resulting truncated Green function.  

The relation to correlation functions in the conformal field theory now
follows trivially.  Witten\Witt\ has shown that the CFT correlators are
given in terms of the truncated bulk correlators and the bulk-boundary
propagator as 
\eqn\cftcorr{\langle T\prod_a \calo(b_a) \rangle = \int \prod_a \left[ dV_a
\gbdel(b_a,x_a)\right] \langle 0|T \prod_a \Phi(x_a) |0\rangle_T\ .}
After substituting \wavepack\ into \lsz, $S_\partial$ is therefore given by
\eqn\btob{S_\partial[f_1\cdots f_m;f_1'\cdots f_n'] =\int \prod_i\left[db_i
{f_i(b_i)\over  Z_i} \right] \int \prod_j\left[db_j'
{f_j'(b_j')\over Z_j'} \right] 
\langle T\prod_i \calo(b_i) \prod_j \calo(b_j')\rangle\
\ ,}
or in the plane wave limit $f\rightarrow e^{-i\omega t} Y_{l\vecm}$, 
\eqn\btobf{S_\partial[\{\omega_i,l_i,\vecm_i\};\{\omega_j',l_j',\vecm_j'\}] =
\langle T\prod_i{1\over Z_i} \calo(\omega_i,l_i,\vecm_i)\prod_j {1\over Z_j'}
\calo(\omega_j',l_j',\vecm_j')\rangle\ .}
These strikingly simple relations tells us that the CFT correlators directly
determine the boundary S-matrix, and thus provide an extremely simple
dictionary between scattering amplitudes in \ads\ space and the 
conformal field theory correlation functions.

\bigskip\bigskip\centerline{{\bf Acknowledgments}}\nobreak

I have greatly benefited from discussions with V. Balasubramanian and
A. Lawrence in the course of writing \BGL.  This work was partially
supported by DOE contract DE-FG-03-91ER40618.

\listrefs

\end